\documentclass[sigplan,10pt,screen,nonacm]{acmart}
\setkeys{acmart.cls}{balance=false}

\microtypesetup{protrusion=false}

\usepackage{amsmath}
\usepackage{algorithmic}
\usepackage{algorithm}
\usepackage{graphicx}
\usepackage{textcomp}
\usepackage{xcolor}
\usepackage{booktabs}
\usepackage{subcaption}
\usepackage{xspace}
\usepackage{listings}
\DeclareTextFontCommand{\texttt}{\normalfont\ttfamily}
\usepackage{multirow}
\usepackage{tikz}
\usepackage{pgfplots}
\pgfplotsset{compat=1.18}
\usetikzlibrary{shapes.geometric, arrows.meta, positioning, calc, patterns}
\AtBeginDocument{\raggedbottom}

\newcommand{\toolname}{InterType\xspace}
\newcommand{\ie}{\emph{i.e.,}\xspace}
\newcommand{\eg}{\emph{e.g.,}\xspace}

\begin{document}

\title[InterType: Neural Binary Type Inference]{InterType: Breaking the Single-Function Barrier in Neural Binary Type Inference}

\author{Shan Wang}
\affiliation{\institution{\resizebox{\hsize}{!}{University of Minnesota Twin Cities}}\city{Minneapolis}\state{MN}\country{USA}}
\email{wang5692@umn.edu}

\author{Tianming Cui}
\affiliation{\institution{\resizebox{\hsize}{!}{University of Minnesota Twin Cities}}\city{Minneapolis}\state{MN}\country{USA}}
\email{cuixx327@umn.edu}

\author{Stephen McCamant}
\affiliation{\institution{\resizebox{\hsize}{!}{University of Minnesota Twin Cities}}\city{Minneapolis}\state{MN}\country{USA}}
\email{smccaman@umn.edu}

\author{Antonia Zhai}
\affiliation{\institution{\resizebox{\hsize}{!}{University of Minnesota Twin Cities}}\city{Minneapolis}\state{MN}\country{USA}}
\email{zhai@umn.edu}

\author{Pen-Chung Yew}
\affiliation{\institution{\resizebox{\hsize}{!}{University of Minnesota Twin Cities}}\city{Minneapolis}\state{MN}\country{USA}}
\email{yew@umn.edu}

\renewcommand{\shortauthors}{Wang et al.}

\begin{abstract}
Variable type recovery is a critical task in binary reverse engineering, enabling downstream applications such as vulnerability discovery, malware analysis, and security enforcement.
While recent Transformer-based models have demonstrated strong performance in type inference by learning instruction-level semantics, they predominantly operate within single-function boundaries (\ie intra-procedural analysis), ignoring the rich type constraints that propagate across function calls.
Empirical studies have shown that nearly 44\% of variables in real-world binaries propagate across function boundaries as arguments or return values, representing a significant source of type information that intra-procedural models cannot exploit.
In this paper, we propose \toolname, a method that embeds inter-procedural analysis into a pre-trained Transformer model for binary type recovery.
Our approach builds on CodeArt, a state-of-the-art Transformer model for stripped binary analysis, and extends it by incorporating caller--callee data flow dependencies during fine-tuning.
Specifically, we extract data flow dependence for each caller--callee pair from disassembled binaries and construct training instances that provide the model with cross-function context.
By exposing the Transformer to inter-procedural variable propagation patterns, our model learns to resolve types that depend on cross-function data flow.
We evaluate \toolname on the coreutils benchmark compiled with optimization levels O0 through O3.
On O1 binaries, \toolname improves recall by 2.76 percentage points, from 89.00\% for the intra-procedural CodeArt baseline to 91.76\%, corresponding to a 25.1\% reduction in recall error rate ($1-\text{recall}$).
\end{abstract}



\maketitle

\section{Introduction}
\label{sec:intro}

Binary reverse engineering is a fundamental activity in software security, enabling analysts to understand the behavior of programs when source code is unavailable.
Among the many tasks in reverse engineering, \emph{variable type recovery}---the process of inferring the original source-level types (\eg \texttt{int}, \texttt{char*}) of variables in compiled, stripped binaries---is particularly important.
Accurate type information dramatically improves the readability of decompiled code, transforming opaque register and memory operations into meaningful, typed expressions that human analysts can reason about.

Type recovery is critical across multiple security-relevant domains.
In \emph{vulnerability discovery}, knowing that a variable is a pointer to a heap-allocated structure rather than a plain integer enables detection of use-after-free and type-confusion vulnerabilities~\cite{lee2011tie}.
In \emph{malware analysis}, recovered types reveal the data structures used by malicious code, such as command-and-control protocol buffers or encryption key schedules, accelerating the understanding of malware behavior.
For \emph{security enforcement mechanisms} such as Control-Flow Integrity (CFI)~\cite{abadi2005cfi}, precise function type signatures are essential for constructing accurate call graphs and validating indirect control transfers.

\paragraph{Limitations of Intra-Procedural Models.}
Recent advances in machine learning have produced increasingly capable models for binary type inference.
Transformer-based approaches such as StateFormer~\cite{pei2021stateformer}, DIRTY~\cite{chen2022dirty}, and CodeArt~\cite{su2024codeart} have demonstrated strong performance by learning instruction-level semantics from large corpora of binary code.
CodeArt, in particular, introduces attention regularization through Masked Dependence Modeling (MDM) to align the model's attention with data-flow dependencies, achieving state-of-the-art results on stripped binary analysis.

However, these models share a fundamental limitation: they operate exclusively within single-function boundaries.
Each function is processed in isolation, and the model has no visibility into how variables are passed to callees as arguments, how return values flow back to callers, or how a variable's type is constrained by its usage in distant functions.
This \emph{intra-procedural} design means that when a variable's type can only be disambiguated through its usage in a callee function---or through a chain of calls---the model lacks the necessary context and defaults to generic type predictions (\eg \texttt{void*} instead of specific pointers).

\paragraph{Motivating Example.}
To illustrate this limitation concretely, consider the \texttt{\_\_stream} variable used internally by the \texttt{do\_decode} function in the \texttt{base32} utility from GNU Coreutils (compiled at optimization level O1).
This variable represents a standard \texttt{FILE} structure stream used for file I/O operations. In the \texttt{base32} binary, the caller function \texttt{do\_decode} maintains a pointer to this stream in the \texttt{r15} register, and passes it to an I/O routine (\eg \texttt{fread\_unlocked}) via the \texttt{rcx} register.

\begin{figure*}[tp]
    \centering
    \begin{minipage}[t]{0.48\textwidth}
        \textbf{Caller (\texttt{do\_decode})\\Intra-Procedural View}
        \begin{lstlisting}[language={[x86masm]Assembler}, basicstyle=\ttfamily\normalsize, frame=single, numbers=none, breaklines=true, breakatwhitespace=true, columns=fullflexible]
do_decode:
    ; ... prepare for file reading ...
    mov  rcx, r15 ;  pass '__stream' to rcx
    mov  edx, 1 ;
    mov  rsi, r14 ;
    call fread_unlocked ;  call I/O routine
    ; ...
        \end{lstlisting}
        \vspace{0.2em}
        \textbf{CodeArt Prediction:} \textcolor{red}{\texttt{base(char)*}} (Incorrect)
    \end{minipage}
    \hfill
    \begin{minipage}[t]{0.48\textwidth}
        \textbf{Callee (\texttt{fread\_unlocked})\\Inter-Procedural View}
        \begin{lstlisting}[language={[x86masm]Assembler}, basicstyle=\ttfamily\normalsize, frame=single, numbers=none, breaklines=true, breakatwhitespace=true, columns=fullflexible]
fread_unlocked:
    ; ... perform stream operations ...
    mov  rax, [rcx] ;  read stream field
    cmp  [rcx+8], rax ;  another stream field
    jle  loc_fill_buffer ;
    mov  [rcx], rax ;
    ; ...
        \end{lstlisting}
        \vspace{0.2em}
        \textbf{\toolname Prediction:} \textcolor{teal}{\texttt{struct*}} (Correct)
    \end{minipage}
    \caption{Side-by-side assembly demonstrating backward type propagation. The intra-procedural model only sees the caller preparing an ambiguous pointer argument, resulting in an incorrect \texttt{base(char)*} prediction. The inter-procedural model incorporates the callee's explicit structural field accesses, correctly deducing that the caller's \texttt{\_\_stream} variable stored in register \texttt{r15} is a \texttt{struct*}.}
    \Description{Two assembly listings compare the caller do_decode with the callee fread_unlocked. The caller copies a pointer from r15 into rcx before the call. The callee accesses fields at offsets zero and eight through rcx. CodeArt predicts a character pointer; InterType uses the callee context to predict a structure pointer.}
    \label{fig:motivating_assembly}
\end{figure*}

As shown in Figure~\ref{fig:motivating_assembly}, the example specifically analyzes the variable \texttt{\_\_stream} stored in the \texttt{r15} register within the \texttt{do\_decode} function of the \texttt{base32} binary. The intra-procedural CodeArt model analyzing the caller only sees an opaque pointer being passed into the \texttt{rcx} argument register. Lacking visibility into the subroutine, CodeArt incorrectly predicts the \texttt{base(char)*} type (a generic byte buffer) for the \texttt{\_\_stream} variable. In contrast, when inter-procedural analysis is applied, \toolname merges the control flow graphs, gaining visibility into how the pointer is used inside the callee. By observing the callee dereferencing the pointer at specific field offsets (\eg \texttt{[rcx]}, \texttt{[rcx+8]}) to access the file stream's internal buffers, \toolname recognizes the signature of a structure. Crucially, this structural type constraint propagates \emph{backward} across the function boundary to the caller, empowering \toolname to correctly predict \texttt{struct*} for the \texttt{\_\_stream} variable in \texttt{r15}. This directly demonstrates that inter-procedural data flow analysis can successfully resolve types where intra-procedural analysis fails.

\paragraph{The Need for Inter-Procedural Analysis.}
The importance of cross-function type propagation is well established.
An empirical study by ByteTR~\cite{li2026bytetr} on the TYDA dataset---comprising over 163,000 binaries across four architectures and four optimization levels---found that \emph{nearly 44\% of variables propagate across function boundaries} as arguments or return values.
This means that intra-procedural models are fundamentally unable to access nearly half of the available type constraints.
The study further reported an 8.46\% improvement in type prediction precision when extending intra-procedural analysis to a two-level direct caller--callee pair, and an additional 2.08\% improvement when extending to three levels. ByteTR counts function levels, so its depth 2 corresponds to one call edge.

While some approaches have attempted to capture inter-procedural information, they rely on Graph Neural Networks (GNNs) to propagate type constraints along variable propagation graphs.
ByteTR~\cite{li2026bytetr} constructs inter-procedural variable propagation graphs and uses Gated Graph Neural Networks (GGNNs) for type prediction.
This approach demonstrates the value of representing cross-function dependencies explicitly.
We investigate how to incorporate such dependencies into the learned assembly representations of a pre-trained Transformer.

\paragraph{LLM-based Methods for Structure Recovery.}
A separate line of work has explored leveraging large language models (LLMs) for binary type recovery, with a particular focus on recovering composite structure types.
ReSym~\cite{xie2024resym} fine-tunes StarCoder to predict variable types and structure field layouts from decompiled code, using a Prolog-based aggregation algorithm to cross-check results across multiple LLM queries.
TypeForge~\cite{wang2025typeforge} constructs Type Flow Graphs from stripped binaries and employs an LLM-based double-elimination framework to select best-fit composite type declarations by comparing the readability of decompiled-code variants.
Both pipelines therefore depend on decompiler output for their LLM-based prediction or selection stages. \toolname instead takes stripped binaries as input and infers types from disassembled instructions and extracted data-flow dependencies, without requiring decompiled code.
ReSym and TypeForge address structure recovery, while our study focuses on token-level classification over a fixed type vocabulary.
We use CodeArt's approximately 109M-parameter pre-trained Transformer to examine the effect of adding inter-procedural context within an existing type-classification pipeline.

\paragraph{Our Approach and Contributions.}
In this paper, we propose \toolname, a method that bridges this gap by embedding inter-procedural analysis directly into a Transformer-based model.
We build on the pre-trained CodeArt model and extend it with inter-procedural context during fine-tuning.
Our key insight is that by merging caller and callee control flow graphs before reaching definition analysis, the analysis naturally discovers data flow paths that cross function boundaries---argument passing and return value propagation.
The resulting inter-procedural dependency edges are encoded as \emph{graph-pattern entries} in CodeArt's \texttt{MaskBuilder}, the same mechanism that encodes intra-procedural data flow.
These graph-pattern entries enable bidirectional attention between caller argument-setup instructions and callee parameter-use instructions, allowing the Transformer's self-attention to propagate type constraints across function boundaries.
Although we evaluate our inter-procedural binary analysis technique only on type recovery, encoding cross-function data flow edges as graph-pattern attention mask entries may also benefit other Transformer-based binary analysis tasks, including function signature recovery, variable name prediction, and vulnerability detection.

Our contributions are as follows:
\begin{enumerate}
    \item We propose \toolname, a method that embeds inter-procedural caller--callee data flow analysis into a pre-trained Transformer model for binary type recovery by merging caller--callee CFGs and running reaching definition analysis on the composite graph.
    \item We show that inter-procedural data flow edges, once extracted from the merged CFG, are directly encoded as graph-pattern entries in the \texttt{MaskBuilder}'s structured attention mask, enabling bidirectional cross-function attention between caller and callee instructions.
    \item We evaluate \toolname on the coreutils benchmark across multiple optimization levels (O0--O3). On O1 binaries, recall improves by 2.76 percentage points, from 89.00\% to 91.76\%, corresponding to a 25.1\% reduction in recall error rate compared to the intra-procedural CodeArt baseline.
    \item We provide interpretability analysis through real-world case studies comparing \toolname with CodeArt, demonstrating the practical benefit of our inter-procedural transformer based approach for type recovery.
\end{enumerate}

\section{Background}
\label{sec:background}

This section provides the necessary background on binary type recovery, intermediate representations used in our approach, and the CodeArt pre-trained model that serves as our foundation.

\subsection{Binary Type Recovery}
\label{sec:bg-type-recovery}

When source code is compiled into a binary executable, the compiler strips away high-level type information.
Variables that were originally declared as \texttt{int}, \texttt{char*}, or \texttt{struct~file\_info*} become undifferentiated values in registers and memory locations.
Stripped binaries---those compiled without debug symbols---present an even greater challenge, as all symbol names, function boundaries, and type annotations are removed.

The goal of \emph{type recovery} is to reconstruct the original source-level types from the compiled binary.
This can be formulated as a classification problem: given a variable (identified by its storage location---a register or memory address---and the instruction context in which it appears), predict its type from a predefined type vocabulary.

\subsection{Ground Truth Extraction from Debug Information}
\label{sec:bg-dwarf}

To train and evaluate type recovery models, we require ground truth type annotations for variables in binary code.
We obtain these from DWARF debug information embedded in non-stripped binaries compiled with the \texttt{-g} flag.
DWARF records contain detailed type information for each variable, including its name, declared type, and \emph{location list}---a mapping from instruction address ranges to storage locations (registers or stack offsets) where the variable resides during execution.

For each variable, we extract its type from the DWARF \texttt{DW\_AT\_type} attribute and its storage locations from the \texttt{DW\_AT\_location} attribute.
This allows us to associate each register or memory reference at each instruction with its ground truth type.
During training, we use the debug-compiled binary for ground truth extraction and the corresponding stripped binary for input features, simulating the real-world scenario where type information is unavailable.

\subsection{CodeArt: Pre-trained Transformer for Stripped Binaries}
\label{sec:bg-codeart}

CodeArt~\cite{su2024codeart} is a BERT-like Transformer model pre-trained on stripped binary code.
It addresses the challenge of learning meaningful representations from assembly code where symbolic information is absent.
Understanding how CodeArt processes data dependencies is essential to appreciating how our inter-procedural extension works, because the same mechanism that encodes intra-procedural data flow can be leveraged to encode cross-function dependencies.

\paragraph{Pre-training Objectives.}
CodeArt employs two pre-training objectives:
\begin{itemize}
    \item \textbf{Masked Language Modeling (MLM):} Following BERT~\cite{devlin2019bert}, random tokens in the assembly code are masked and the model is trained to predict them, learning instruction-level semantics.
    \item \textbf{Masked Dependence Modeling (MDM):} A novel objective that regularizes the model's attention to align with data-flow dependencies extracted via program analysis. This encourages the model to attend to semantically related tokens (\eg a register definition and its subsequent uses) rather than relying solely on positional proximity.
\end{itemize}

\paragraph{Data Dependency Extraction.}
For each function, CodeArt performs \emph{reaching definition analysis} on the function's control flow graph (CFG) to extract data dependencies.
Each instruction is analyzed for its \emph{definitions} (registers/memory locations written) and \emph{uses} (registers/memory locations read).
For each use of a variable at instruction $i_k$, the analysis identifies the set of instructions that may have last defined that variable---either within the same basic block (\emph{intra-block} dependencies) or across basic blocks via the reaching definition fixpoint computation (\emph{inter-block} dependencies).
The resulting dependencies are represented as a set of directed edges $D = \{(i_k, i_j) \mid i_k \text{ uses a value defined by } i_j\}$.

\paragraph{Structured Attention Masking.}
CodeArt does not use a standard full attention mask.
Instead, it employs a \texttt{MaskBuilder} that constructs a \emph{structured attention mask} $\mathbf{M} \in \{0,1\}^{L \times L}$ from four composable patterns:
\begin{enumerate}
    \item \textbf{Global memory pattern:} The \texttt{[CLS]} token attends to all tokens, and all tokens attend to \texttt{[CLS]}, providing a global information aggregation point.
    \item \textbf{Local pattern:} Tokens within the same instruction attend to each other. Each instruction is delimited by a special \texttt{<INST>} bridge token.
    \item \textbf{Bridge pattern:} All tokens can attend to the \texttt{<INST>} bridge tokens of all instructions, enabling coarse-grained cross-instruction attention.
    \item \textbf{Graph pattern:} For each data dependency edge $(i_k, i_j) \in D$, bidirectional attention is enabled between the bridge tokens of instructions $i_k$ and $i_j$: $\mathbf{M}[\text{pos}(i_k), \text{pos}(i_j)] = \mathbf{M}[\text{pos}(i_j), \text{pos}(i_k)] = 1$.
\end{enumerate}
The graph pattern is the critical channel through which data flow semantics are injected into the Transformer's self-attention.
In addition, CodeArt computes a \emph{relative position matrix} from the dependency graph, encoding the shortest-path distance between instructions through the transitive closure of data dependencies.
This provides the model with a notion of how ``far apart'' two instructions are in the data flow graph, complementing the absolute positional encoding.

\paragraph{Data Collation Pipeline.}
During fine-tuning for type inference, the data collator processes each training example by: (1)~tokenizing the assembly instruction sequence and dependency edges via \texttt{inst\_encode}, which invokes the \texttt{MaskBuilder} to produce the structured attention mask and relative position matrix; (2)~aligning ground truth type labels to individual tokens based on register or memory operand matching via DWARF debug information; and (3)~padding/truncating to the model's maximum sequence length (512 tokens).

\paragraph{Limitation.}
In CodeArt's original pipeline, each function's CFG is processed independently: the reaching definition analysis operates on a single function's CFG, and the resulting data dependencies are strictly intra-procedural.
The \texttt{MaskBuilder} constructs an attention mask that only permits graph-pattern attention between instructions \emph{within the same function}.
Consequently, when a variable's type is determined by its usage in a callee function, no dependency edge exists to guide the model's attention across the function boundary, and the model lacks the context to infer the correct type.
Our work extends this pipeline by modifying the CFG construction and dependency extraction stages to produce \emph{inter-procedural} data dependency edges.
These cross-boundary edges are encoded as graph-pattern entries in the \texttt{MaskBuilder}'s structured attention mask, enabling the Transformer's self-attention to propagate type constraints between caller and callee instructions.

\section{Binary Type Recovery through Embedded Inter-Procedural Analysis}
\label{sec:methodology}

This section presents \toolname, our approach to embedding inter-procedural analysis in a Transformer-based model for binary type recovery.
The key insight is that inter-procedural data flow edges---extracted by merging caller and callee CFGs and running reaching definition analysis on the composite graph---can be encoded as \emph{graph-pattern entries} in CodeArt's \texttt{MaskBuilder}.
These entries enable bidirectional attention between instructions across the function boundary, allowing the Transformer to propagate type constraints from callee field accesses back to caller argument setup and vice versa.

\subsection{Overview}
\label{sec:method-overview}

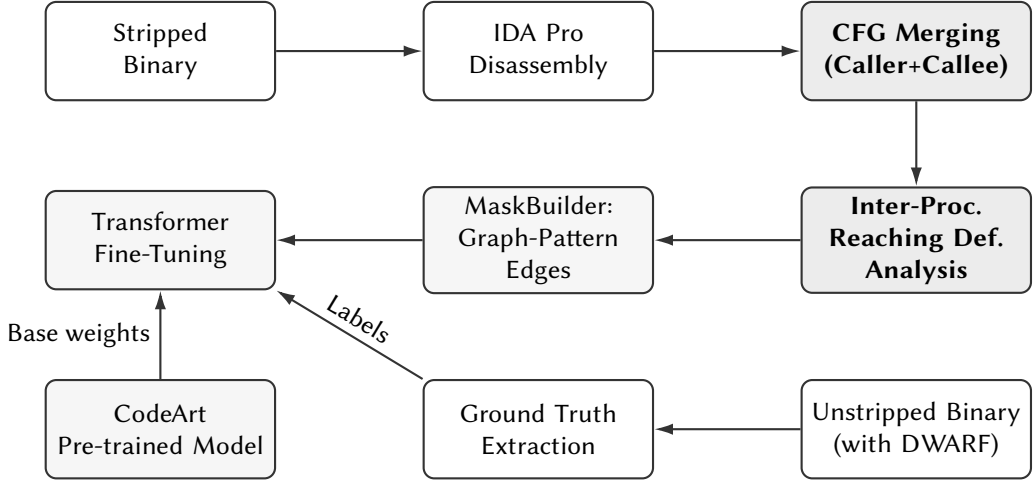
\begin{figure*}[tp]
    \centering
    \begin{tikzpicture}[
        block/.style={rectangle, draw=black!80, thick, fill=white, text width=2.8cm, align=center, rounded corners, minimum height=1.3cm, font=\sffamily\normalsize},
        highlight/.style={block, fill=gray!15, font=\sffamily\normalsize\bfseries},
        model/.style={block, fill=gray!7},
        line/.style={draw, thick, -{Latex[length=2.5mm, width=1.5mm]}, draw=black!80}
    ]
    \node [block] (stripped) at (0,0) {Stripped\\Binary};
    \node [block] (ida) at (5,0) {IDA Pro\\Disassembly};
    \node [highlight] (cfg) at (10,0) {CFG Merging\\(Caller+Callee)};
    \node [highlight] (rda) at (10,-2.5) {Inter-Proc.\\Reaching Def.\\Analysis};
    \node [model] (mask) at (5,-2.5) {MaskBuilder:\\Graph-Pattern\\Edges};
    \node [model] (finetune) at (0,-2.5) {Transformer\\Fine-Tuning};
    \node [block] (unstripped) at (10,-5) {Unstripped Binary\\(with DWARF)};
    \node [block] (dwarf) at (5,-5) {Ground Truth\\Extraction};
    \node [model] (codeart) at (0,-5) {CodeArt\\Pre-trained Model};
    \path [line] (stripped) -- (ida);
    \path [line] (ida) -- (cfg);
    \path [line] (cfg) -- (rda);
    \path [line] (rda) -- (mask);
    \path [line] (mask) -- (finetune);
    \path [line] (unstripped) -- (dwarf);
    \path [line] (dwarf.north west) -- (finetune.south east)
        node[midway, above, sloped, font=\sffamily\normalsize] {Labels};
    \path [line] (codeart) -- (finetune)
        node[midway, left, font=\sffamily\normalsize] {Base weights};
    \end{tikzpicture}
    \caption{Overview of the \toolname framework. The pipeline merges caller--callee CFGs and runs reaching definition analysis to produce inter-procedural data dependencies, which are injected into CodeArt's structured attention mask as graph-pattern edges.}
    \Description{A flow diagram shows stripped binaries passing through IDA Pro disassembly, caller-callee CFG merging, inter-procedural reaching definition analysis, and MaskBuilder to Transformer fine-tuning. A separate unstripped binary supplies DWARF ground-truth labels, and the pretrained CodeArt model supplies initial weights.}
    \label{fig:framework}
\end{figure*}

Figure~\ref{fig:framework} illustrates the overall framework of \toolname.
Our approach consists of four main steps:

\begin{enumerate}
    \item \textbf{Pre-trained Model:} We adopt the CodeArt~\cite{su2024codeart} pre-trained Transformer (approximately 109M parameters) as our foundation, leveraging its learned representations of assembly code semantics.
    \item \textbf{Ground Truth Extraction:} We compile benchmark programs with debug information (\texttt{-g} flag) and extract ground truth types from DWARF debug entries, mapping each variable to its type and storage locations across registers and memory.
    \item \textbf{Inter-Procedural Data Flow Extraction:} We disassemble the corresponding stripped binary using IDA Pro~\cite{hex2026idapro}, extract the CFGs of both caller and callee for each call site, merge them into a single composite CFG, and run reaching definition analysis on the merged graph to produce both intra- and inter-procedural data dependency edges.
    \item \textbf{Fine-tuning with Inter-Procedural Context:} The merged instruction sequence and augmented data dependencies are fed through the CodeArt tokenizer, \texttt{MaskBuilder}, and data collator pipeline. The \texttt{MaskBuilder} encodes inter-procedural dependency edges as graph-pattern entries in the structured attention mask, enabling bidirectional attention between caller argument-setup instructions and callee parameter-use instructions across the function boundary.
\end{enumerate}

\subsection{Intra-Procedural Baseline: Original CodeArt Pipeline}
\label{sec:method-intra}

In the original CodeArt pipeline, each function is processed independently through the following stages:

\paragraph{Step 1: CFG Extraction.}
IDA Pro disassembles the stripped binary and extracts the CFG for each function as a directed graph $G_f = (B_f, E_f)$, where $B_f$ is the set of basic blocks and $E_f$ is the set of control flow edges.
Each basic block contains a sequence of assembly instructions with their addresses.

\paragraph{Step 2: Data Dependency Analysis.}
The original CodeArt pipeline takes the single-function CFG and runs data flow analysis, which performs:
\begin{enumerate}
    \item \textbf{Reaching Definition Analysis:} A standard fixpoint computation on the CFG that determines, for each variable use at instruction $i_k$, the set of instructions that may have most recently defined that variable. The analysis maintains, for each basic block, an \texttt{IN} set of reaching definitions and propagates them through the CFG until convergence.
    \item \textbf{Dependency Edge Extraction:} For each instruction $i_k$ and each variable it uses, the analyzer looks up the reaching definition---first in intra-block definitions (from the same basic block), then in inter-block definitions (from \texttt{bb\_in} of the enclosing basic block). Each match produces a dependency edge $(i_k, i_j)$.
\end{enumerate}
Since the CFG contains only a single function, all resulting dependency edges are intra-procedural. The output is a list of \texttt{(code, data\_dep)} pairs, where \texttt{data\_dep} is a list of instruction ID pairs $[(i_k, i_j), \ldots]$.

\paragraph{Step 3: Attention Mask Construction.}
CodeArt passes each example's \texttt{code} and \texttt{data\_dep} to the tokenizer's instruction encoding method, which invokes the \texttt{MaskBuilder}.
For each dependency edge $(i_k, i_j) \in \texttt{data\_dep}$, the \texttt{MaskBuilder} enables bidirectional attention between the bridge tokens of instructions $i_k$ and $i_j$ in the attention mask.
All dependency edges are between instructions within the same function, so the graph-pattern attention is strictly intra-procedural.

\subsection{Inter-Procedural Data Flow Extraction}
\label{sec:method-inter-extraction}

\toolname builds the program analysis pipeline that produces \emph{inter-procedural} data dependency edges, including two stages: CFG construction and dependency analysis.

\paragraph{Step 1: Caller--Callee CFG Collection.}
For each direct call instruction in the binary, we identify the call site address and the target callee function.
We extract the CFGs of both the caller and callee as separate directed graphs $G_{\text{caller}} = (B_c, E_c)$ and $G_{\text{callee}} = (B_e, E_e)$.
Both CFGs are stored as a tuple \texttt{(caller\_cfg, callee\_cfg)}, along with the call site address.

\paragraph{Step 2: CFG Merging.}
In our modified data flow analysis, when a caller--callee pair is detected, we merge the two CFGs into a single composite graph:
\begin{equation}
\begin{aligned}
G_{\text{merged}} ={}& G_{\text{caller}} \cup G_{\text{callee}} \\
&{}\cup \{(b_{\text{call}}, b_{\text{entry}}), (b_{\text{exit}}, b_{\text{ret}})\}.
\end{aligned}
\end{equation}
where $b_{\text{call}}$ is the caller's basic block containing the call instruction, $b_{\text{entry}}$ is the callee's entry block (the node with in-degree 0 in $G_{\text{callee}}$), $b_{\text{exit}}$ is the callee's exit block (nodes with out-degree 0), and $b_{\text{ret}}$ is the caller's basic block that follows the call site (the return point).
The edge $(b_{\text{call}}, b_{\text{entry}})$ models the call transfer, and edges $(b_{\text{exit}}, b_{\text{ret}})$ model the return.

\paragraph{Step 3: Reaching Definition Analysis on the Merged CFG.}
The merged CFG is passed to the data flow analyzer that CodeArt uses for single functions.
Because the caller and callee CFGs are now connected, the reaching definition analysis naturally propagates definitions across the function boundary:

\begin{itemize}
    \item \textbf{Argument passing (caller $\to$ callee):}
    When the caller writes a value to an argument register (\eg \texttt{mov rdi, rax}) before the call instruction, this definition reaches the callee's entry block through the $(b_{\text{call}}, b_{\text{entry}})$ edge.
    When the callee's first instruction uses \texttt{rdi}, the reaching definition analysis identifies the caller's \texttt{mov} instruction as the definition, producing an inter-procedural dependency edge.
    
    \item \textbf{Return value propagation (callee $\to$ caller):}
    When the callee writes to \texttt{rax} before returning, this definition reaches the caller's return-site block through the $(b_{\text{exit}}, b_{\text{ret}})$ edges.
    When the caller subsequently uses \texttt{rax}, the analysis produces an inter-procedural edge from the callee's instruction defining the returned value to the caller's use.
\end{itemize}

\begin{table*}[tp]
\centering
\caption{Pipeline comparison: original CodeArt (intra-procedural) vs.\ \toolname (inter-procedural).}
\label{tab:pipeline-diff}
\normalsize
\begin{tabular}{p{0.16\textwidth}p{0.34\textwidth}p{0.42\textwidth}}
\toprule
\textbf{Component} & \textbf{CodeArt} & \textbf{\toolname} \\
\midrule
CFG input & Single function CFG & Merged caller--callee CFG with call/return edges \\
\midrule
Reaching def. analysis & Intra-procedural only & On merged CFG; produces cross-boundary edges \\
\midrule
Data dependency & Intra-procedural edges only & Unified intra + inter-procedural edges \\
\midrule
Attention mask & Graph-pattern entries within one function & Graph-pattern entries across caller--callee boundary \\
\midrule
Transformer model & CodeArt ($\sim$109M parameters) & CodeArt ($\sim$109M parameters) \\
\midrule
Training config & LR=$10^{-4}$, 5 epochs, batch=48 & LR=$10^{-4}$, 5 epochs, batch=48 \\
\bottomrule
\end{tabular}
\end{table*}

\subsection{Fine-tuning with Inter-Procedural Context}
\label{sec:method-finetune}

The inter-procedural dependency edges produced by the merged CFG analysis are encoded as graph-pattern entries in the \texttt{MaskBuilder}'s structured attention mask, enabling the Transformer's self-attention to attend across the caller--callee boundary.

\paragraph{Input Sequence Construction.}
The merged instruction sequence contains both caller and callee instructions in basic-block order:
\begin{equation}
X_{\text{pair}} = [\texttt{[CLS]}, \underbrace{x_1^c, \ldots, x_p^c}_{\text{caller instrs}}, \underbrace{x_1^e, \ldots, x_q^e}_{\text{callee instrs}}, \texttt{[SEP]}]
\end{equation}
Each instruction is preceded by an \texttt{<INST>} bridge token.

\paragraph{Attention Mask with Cross-Function Edges.}
The \texttt{MaskBuilder} processes the unified \texttt{data\_dep} list, which now includes inter-procedural edges.
For each inter-procedural edge $(i_k^e, i_j^c)$ where $i_k^e$ is a callee instruction and $i_j^c$ is a caller instruction, the \texttt{MaskBuilder} sets:
\begin{equation}
\mathbf{M}[\text{pos}(i_k^e), \text{pos}(i_j^c)] = \mathbf{M}[\text{pos}(i_j^c), \text{pos}(i_k^e)] = 1
\end{equation}
This enables \emph{bidirectional attention} between the caller's argument-setup instruction and the callee's parameter-use instruction, and between the callee's instruction defining the returned value and the caller's return-value use.
The relative position matrix similarly computes shortest-path distances through the transitive closure of the combined dependency graph, allowing the model to learn that an inter-procedural edge represents a ``close'' semantic relationship despite potentially large positional distance in the token sequence.

\paragraph{Type Label Assignment.}
The data collator assigns ground truth type labels to tokens in both the caller and callee portions of the sequence.
For each token corresponding to a register or memory operand, the collator looks up the type from the DWARF-derived type annotations, matching by register name or memory operand.
The register-to-type mapping handles aliased registers (\eg \texttt{eax}, \texttt{al} $\to$ \texttt{rax}) to ensure consistent label assignment.
Tokens that do not correspond to typed variables receive the \texttt{noacc} (``no access'') label and are ignored during loss computation (label = $-100$).

\paragraph{Training Objective.}
The model is trained with a standard cross-entropy loss over the type vocabulary:
\begin{equation}
\mathcal{L} = -\sum_{j \in \mathcal{V}} \log P(y_j | x_j, X_{\text{pair}}, \mathbf{M})
\end{equation}
where $\mathcal{V}$ is the set of token positions with valid type labels (both caller and callee), $y_j$ is the ground truth type, and $\mathbf{M}$ is the structured attention mask containing both intra- and inter-procedural graph-pattern edges.

\paragraph{Summary of Differences from Original CodeArt.}
Table~\ref{tab:pipeline-diff} summarizes the key differences between the original CodeArt pipeline and \toolname.
The core change is that inter-procedural data dependency edges---extracted from the merged caller--callee CFG---are encoded as additional graph-pattern entries in the \texttt{MaskBuilder}'s structured attention mask, extending the Transformer's attention from intra-procedural to cross-function scope.

\section{Evaluation}
\label{sec:evaluation}

We evaluate \toolname to answer the following research questions:

\begin{itemize}
    \item \textbf{RQ1 (Accuracy):} How does embedding inter-procedural analysis improve type recovery accuracy compared to the intra-procedural baseline?
    \item \textbf{RQ2 (Optimization Levels):} How robust is \toolname across different compiler optimization levels (O0--O3)?
    \item \textbf{RQ3 (Scalability):} Does \toolname generalize across different benchmark suites?
    \item \textbf{RQ4 (Interpretability):} Can we demonstrate concrete cases where inter-procedural context enables correct type recovery?
\end{itemize}

\subsection{Experimental Setup}
\label{sec:eval-setup}

\paragraph{Dataset.}
We use the GNU Coreutils benchmark (version 8.30) as our primary evaluation dataset.
Coreutils is a standard benchmark in binary analysis research, containing over 100 utility programs with diverse data types and calling patterns.
We compile each program using GCC with four optimization levels: \texttt{-O0}, \texttt{-O1}, \texttt{-O2}, and \texttt{-O3}.
The 106 Coreutils O1 binaries are 64-bit little-endian x86-64 executables built with GCC 5.4.0, as verified from their ELF headers and compiler metadata.
Their DWARF producer records list the compilation flags \texttt{-mtune=generic}, \texttt{-march=x86-64}, \texttt{-g}, \texttt{-O1}, and \texttt{-fstack-protector-strong}.
Each binary is compiled with debug information (\texttt{-g}) for ground truth extraction, and a stripped copy is generated for model input.
We split the dataset into training (70\%), validation (15\%), and test (15\%) sets at the binary level, ensuring that no binary appears in more than one split to prevent data leakage.
We additionally evaluate on a mix of Coreutils, Diffutils, and Findutils to assess cross-dataset generalization.

\paragraph{Baselines.}
We compare \toolname against the following baselines:
\begin{itemize}
    \item \textbf{CodeArt}~\cite{su2024codeart}: The intra-procedural Transformer baseline, using the same pre-trained model without inter-procedural context.
    \item \textbf{ByteTR}~\cite{li2026bytetr}: An inter-procedural GNN-based approach that constructs variable propagation graphs across caller--callee boundaries. For Table~\ref{tab:main-results}, we evaluate the authors' released checkpoint trained on TYDA on our binary-level Coreutils test split, without retraining it on our Coreutils training split.
\end{itemize}

\paragraph{ByteTR Evaluation Protocols.}
We distinguish the published protocol, the released implementation examined in our reproduction, and the settings used for our results.
ByteTR's published evaluation describes a project-level separation between training and test data~\cite[Section~5.1.1]{li2026bytetr}.
In the released implementation we examined, samples were randomly split at the variable level; the authors confirmed this implementation behavior in email correspondence.
This observation concerns that released implementation and should not be read as a description of the protocol stated in the paper.
Our main comparison uses the released TYDA checkpoint, whereas CodeArt and \toolname are fine-tuned on our training data.
Separately, we retrained ByteTR on a TYDA subset to examine split sensitivity: O1 accuracy was 37.17\% with a binary-level split and 72.37\% with a variable-level split.
These retraining results are separate from the released-checkpoint results in Table~\ref{tab:main-results}.

\paragraph{Metrics.}
We report token-level micro-averaged precision, recall, and F1 over non-\texttt{noacc} type classes.
Predictions of a type at a \texttt{noacc} position count as false positives, while predictions of \texttt{noacc} at a typed position count as false negatives.
We evaluate using standard classification metrics:
\begin{itemize}
    \item \textbf{Precision}: The fraction of predicted types that are correct.
    \item \textbf{Recall}: The fraction of ground truth types that are correctly recovered.
    \item \textbf{F1 Score}: The harmonic mean of precision and recall.
\end{itemize}

We also report \emph{error rate reduction}, defined as $1 - \frac{1 - R_{\text{ours}}}{1 - R_{\text{baseline}}}$, where $R$ denotes recall.
We report absolute recall differences in percentage points alongside the relative reduction in $1-R$.

\paragraph{Implementation Details.}
We fine-tune the pre-trained CodeArt model with a learning rate of $1 \times 10^{-4}$ for 5 epochs.
Experiments in this study are conducted on a server equipped with a AMD EPYC 7763 64-Core Processor and NVIDIA A100 GPU running the Rocky Linux 8.10 operating system.
We limit the inter-procedural depth to direct caller--callee pairs (one call edge, termed depth 1 in this paper). This corresponds to ByteTR's two-level setting~\cite{li2026bytetr}.

\subsection{RQ1: Accuracy}
\label{sec:eval-rq1}

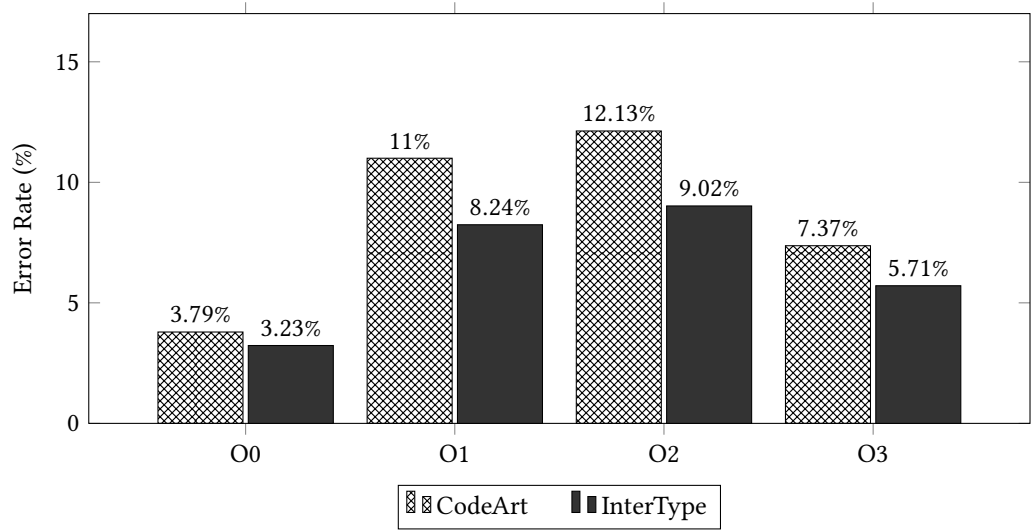
\begin{figure*}[tp]
    \centering
    \begin{tikzpicture}
        \begin{axis}[
            ybar,
            bar width=32pt,
            width=0.85\textwidth,
            height=7cm,
            enlarge x limits=0.25,
            symbolic x coords={O0, O1, O2, O3},
            xtick=data,
            ymin=0, ymax=17,
            ylabel={Error Rate (\%)},
            nodes near coords={\pgfmathprintnumber{\pgfplotspointmeta}\%},
            every node near coord/.append style={font=\normalsize, /pgf/number format/fixed, /pgf/number format/precision=2},
            legend style={at={(0.5,-0.15)}, anchor=north, legend columns=-1, font=\normalsize, /tikz/every even column/.append style={column sep=0.5cm}},
        ]
        \addplot[pattern=crosshatch, pattern color=black, draw=black] coordinates {(O0, 3.79) (O1, 11.00) (O2, 12.13) (O3, 7.37)};
        \addplot[fill=black!80, draw=black] coordinates {(O0, 3.23) (O1, 8.24) (O2, 9.02) (O3, 5.71)};
        \legend{CodeArt, \toolname}
        \end{axis}
    \end{tikzpicture}
    \caption{Recall error rate ($1 - \text{Recall}$) across optimization levels. At \texttt{-O1}, the 2.76-percentage-point recall improvement corresponds to a 25.1\% reduction in recall error rate.}
    \Description{A grouped bar chart compares recall error rates for CodeArt and InterType at four optimization levels. The respective rates are 3.79 and 3.23 percent at O0, 11.00 and 8.24 percent at O1, 12.13 and 9.02 percent at O2, and 7.37 and 5.71 percent at O3. InterType has a lower error rate at every level.}
    \label{fig:error-rate}
\end{figure*}

Table~\ref{tab:main-results} presents the detailed experimental results on the coreutils benchmark across all four optimization levels. \toolname consistently improves precision, recall, and F1 over the intra-procedural CodeArt baseline. The table also reports results for ByteTR's released TYDA checkpoint under the different training conditions described in Section~\ref{sec:eval-setup}. Figure~\ref{fig:error-rate} plots the recall error rate of \toolname compared to CodeArt.

\begin{table*}[tp]
\centering
\caption{Type recovery results on coreutils (O0--O3). CodeArt and \toolname are fine-tuned on our training split. ByteTR rows use the released TYDA checkpoint without retraining on Coreutils; these rows therefore reflect different training conditions.}
\label{tab:main-results}
\begin{tabular}{llccc}
\toprule
\textbf{Optimization} & \textbf{Method} & \textbf{Recall} & \textbf{Precision} & \textbf{F1} \\
\midrule
\multirow{3}{*}{O0} & CodeArt & 0.9621 & 0.9609 & 0.9615 \\
& ByteTR & 0.3378 & 0.3365 & 0.3371 \\
& \textbf{\toolname (ours)} & \textbf{0.9677} & \textbf{0.9664} & \textbf{0.9671} \\
\midrule
\multirow{3}{*}{O1} & CodeArt & 0.8900 & 0.8792 & 0.8845 \\
& ByteTR & 0.2915 & 0.2928 & 0.2921 \\
& \textbf{\toolname (ours)} & \textbf{0.9176} & \textbf{0.9151} & \textbf{0.9163} \\
\midrule
\multirow{3}{*}{O2} & CodeArt & 0.8787 & 0.8536 & 0.8659 \\
& ByteTR & 0.2577 & 0.2567 & 0.2572 \\
& \textbf{\toolname (ours)} & \textbf{0.9098} & \textbf{0.8835} & \textbf{0.8964} \\
\midrule
\multirow{3}{*}{O3} & CodeArt & 0.9263 & 0.9194 & 0.9228 \\
& ByteTR & 0.2133 & 0.2213 & 0.2172 \\
& \textbf{\toolname (ours)} & \textbf{0.9429} & \textbf{0.9329} & \textbf{0.9379} \\
\bottomrule
\end{tabular}
\end{table*}

At \texttt{-O1}, \toolname improves token-level recall by 2.76 percentage points, from 89.00\% for CodeArt to 91.76\%.
This corresponds to a 25.1\% reduction in recall error rate ($1-R$), from 11.00\% to 8.24\%.

The improvement is particularly pronounced for complex context-dependent types such as \emph{pointers} and \emph{structures}. To understand the source of this improvement, we analyzed the detailed confusion matrices of InterType and CodeArt. We found that the accuracy of predicting \texttt{short unsigned int} increased the most, jumping from a complete failure (0.0\%) in CodeArt to 91.0\% in \toolname. Furthermore, \toolname successfully rescued deeply nested types: the accuracy for \texttt{struct**} (double pointers to structures) rose from 0.0\% to 86.4\%, and function pointers (\texttt{subroutine*}) saw a massive 16.5 percentage point increase (from 78.8\% to 95.3\%).

These gains support the benefit of inter-procedural context. Figure~\ref{fig:motivating_assembly} illustrates this benefit: the caller \texttt{do\_decode} passes the \texttt{\_\_stream} pointer from \texttt{r15} to \texttt{rcx} using \texttt{mov rcx, r15} before calling \texttt{fread\_unlocked}. CodeArt sees only the caller and incorrectly predicts \texttt{base(char)*}. By incorporating the callee's accesses to \texttt{[rcx]} and \texttt{[rcx+8]}, \toolname observes how the pointer is used to access structure fields and correctly predicts \texttt{struct*} for the caller's \texttt{\_\_stream} variable.

ByteTR's released TYDA checkpoint yields lower scores on our Coreutils test split than the fine-tuned CodeArt and \toolname models. Because these models differ in training data and adaptation to the evaluation benchmark, this comparison does not isolate the effect of model architecture or data splitting. The separate TYDA-subset reproduction in Section~\ref{sec:eval-setup} shows sensitivity to the split protocol, but does not establish the cause of the Coreutils performance gap. We therefore base our assessment of the inter-procedural extension primarily on the comparison with CodeArt under the shared fine-tuning setup.

\subsection{RQ2: Robustness Across Optimization Levels}
\label{sec:eval-rq2}

Table~\ref{tab:main-results} and Figure~\ref{fig:error-rate} illustrate the performance robustness of \toolname against increasingly aggressive compiler optimizations.

Compiler optimizations significantly affect the difficulty of type recovery.
At \texttt{-O0}, code structure closely mirrors the source, with variables predominantly stored on the stack, allowing both CodeArt and \toolname to achieve F1 scores exceeding 0.96.
However, at higher optimization levels (\texttt{-O1} through \texttt{-O3}), aggressive register allocation, code motion, and structural transformations heavily obscure data flows.
In particular, variables are increasingly stored in registers, making local inference much more challenging.

As clearly shown by the consistent error rate reductions in Figure~\ref{fig:error-rate}, \toolname effectively bridges this gap and maintains a robust advantage over the intra-procedural baseline across every optimization level.
The inter-procedural context provided by our approach is especially valuable at \texttt{-O1} and \texttt{-O2}, where the compiler obscures intra-procedural data flows but retains most function call boundaries. For instance, at \texttt{-O2}, \toolname reduces the recall error rate from 12.13\% (CodeArt) to 9.02\%, demonstrating that cross-function attention effectively recovers the type constraints that local compiler optimizations destroy.

\subsection{RQ3: Scalability Across Software Projects}
\label{sec:eval-rq3}

To assess whether \toolname generalizes beyond a single software project, we evaluated it on a mixed benchmark dataset consisting of \texttt{coreutils-8.30}, \texttt{diffutils-3.4}, and \texttt{findutils-4.6} compiled at the \texttt{-O1} optimization level. This diverse mixture introduces a wider variety of programming idioms, proprietary data structures, and cross-module caller--callee patterns.

\begin{table}[t]
\centering
\caption{Type recovery results on the mixed benchmark (Coreutils, Diffutils, Findutils) at optimization level O1.}
\label{tab:scalability}
\setlength{\tabcolsep}{4pt}
\begin{tabular}{@{}lccc@{}}
\toprule
\textbf{Method} & \textbf{Recall} & \textbf{Precision} & \textbf{F1} \\
\midrule
CodeArt (intra) & 0.8538 & 0.8310 & 0.8423 \\
\textbf{\toolname (ours)} & \textbf{0.8778} & \textbf{0.8638} & \textbf{0.8707} \\
\bottomrule
\end{tabular}
\end{table}

Table~\ref{tab:scalability} presents the results on this generalized dataset. \toolname successfully scales to this more complex, mixed environment, achieving an F1 score of 0.8707 compared to the intra-procedural CodeArt baseline's 0.8423. The consistent margin of improvement across precision, recall, and F1 demonstrates that the inter-procedural context captured by \toolname is robustly applicable across different codebases and does not overfit to the idiosyncratic call graphs of a single repository.

\subsection{RQ4: Interpretability}
\label{sec:eval-rq4}

To demonstrate the practical interpretability and the concrete benefits of inter-procedural analysis, we revisit the case study presented earlier in Figure~\ref{fig:motivating_assembly}. This example directly compares the predictions of the intra-procedural CodeArt baseline against \toolname for the \texttt{\_\_stream} variable stored in the \texttt{r15} register within the \texttt{do\_decode} function (from the \texttt{base32} coreutils binary).

When operating strictly intra-procedurally, CodeArt analyzes the caller (\texttt{do\_decode}) and only observes an opaque pointer being set up for the \texttt{rcx} argument register via \texttt{mov rcx, r15}. Lacking visibility into the subroutine, CodeArt misses the structural cues necessary to deduce the pointer's true purpose, resulting in an incorrect prediction of \texttt{base(char)*} (a generic byte buffer).

In contrast, \toolname seamlessly integrates the control flow graph of the callee (\texttt{fread\_unlocked}) into the caller's context. By tracing the data flow across the function boundary, \toolname's cross-function attention mechanism explicitly observes the callee dereferencing the pointer at specific field offsets (\eg \texttt{[rcx]} and \texttt{[rcx+8]}). These distinct memory access patterns are the signature of a structure field lookup. By successfully propagating this critical structural constraint \emph{backward} from the callee to the caller, \toolname correctly predicts the complex \texttt{struct*} type for the \texttt{\_\_stream} variable in \texttt{r15}. 

This concrete example perfectly illustrates how \toolname provides interpretable, localized type constraints that resolve the inherent ambiguity that plagues intra-procedural models.

\section{Related Work}
\label{sec:related}

Binary type recovery has been approached through static analysis, dynamic analysis, and machine learning methods.
We survey each category and position our work relative to the state of the art.

\subsection{Static Analysis Methods}
\label{sec:related-static}

Traditional static analysis approaches formulate type recovery as a constraint satisfaction problem.
TIE~\cite{lee2011tie} pioneered this approach by generating type constraints from binary instructions and solving them using lattice-based inference.
While principled, TIE's handcrafted rules struggle with complex pointer arithmetic and produce overly conservative types.
SecondWrite~\cite{elwazeer2013secondwrite} improved scalability by lifting binaries to LLVM IR and performing type recovery at the IR level, but remains limited by the precision of the lifting process.
Retypd~\cite{noonan2016retyped} introduced polymorphic type inference to handle subtyping relationships, achieving more precise results for programs with complex type hierarchies.

Industry tools such as IDA Pro~\cite{hex2026idapro} and Ghidra~\cite{ghidra2026} implement heuristic-based type propagation within their decompilation pipelines.
While effective for common patterns, these tools rely on predefined rules and struggle with novel code patterns and aggressive compiler optimizations.

\subsection{Dynamic Analysis Methods}
\label{sec:related-dynamic}

Dynamic analysis approaches infer types from runtime execution traces.
REWARDS~\cite{lin2010rewards} propagates type information from known API calls backward through execution traces, leveraging the well-typed interfaces of system calls and library functions.
Howard~\cite{slowinska2011howard} extends this approach by tracking data structure layouts through memory access patterns during execution.

While dynamic approaches can be highly precise for exercised code paths, they are fundamentally limited by code coverage: only code executed during the trace is analyzed, leaving unexercised paths without type information.
Additionally, the overhead of dynamic instrumentation can be substantial, making these approaches impractical for large-scale analysis.

\subsection{Machine Learning Methods}
\label{sec:related-ml}

Machine learning approaches to binary type recovery span several paradigms:

\paragraph{Classical Models.}
EKLAVYA~\cite{chua2017eklavya} was among the first to apply neural networks (RNN) to predict function argument types.
DEBIN~\cite{he2018debin} uses Extremely Randomized Trees for variable identification, followed by Conditional Random Fields for structured prediction of variable names and types.
TypeMiner~\cite{maier2019typeminer} and BITY~\cite{xu2017bity} employed classical classifiers (random forests, SVMs) on hand-engineered features.
CATI~\cite{chen2020cati} applied convolutional neural networks to capture local instruction patterns.
These approaches are limited by their reliance on hand-engineered features or architectures that cannot capture long-range dependencies.

\paragraph{Graph Neural Networks.}
GNN-based approaches model the program's control and data flow as graphs.
TYGR~\cite{zhu2024tygr} constructs per-function data flow graphs and uses GNNs to infer types for registers and memory locations. Its data-flow representation is intra-procedural.
DRAGON~\cite{stewart2025dragon} extends GNN-based type prediction with learned confidence estimates.
ByteTR~\cite{li2026bytetr} is the most closely related GNN-based work, performing inter-procedural analysis by constructing variable propagation graphs that span caller--callee boundaries and using a GGNN for type prediction.
ByteTR demonstrates the value of inter-procedural analysis using a GGNN over variable propagation graphs; \toolname instead encodes cross-function dependencies in the structured attention of a pre-trained Transformer.

\paragraph{Transformer-based Models.}
StateFormer~\cite{pei2021stateformer} uses a Transformer pre-trained with Generative State Modeling (GSM) to learn operational semantics of assembly code, achieving strong results on type inference.
DIRTY~\cite{chen2022dirty} applies a Transformer encoder-decoder architecture to decompiler pseudo-code for joint variable name and type prediction.
CodeArt~\cite{su2024codeart} introduces attention regularization through Masked Dependence Modeling to align the model's attention with program-analysis-derived data flow for intra-procedural type recovery from stripped binaries.
All three operate exclusively within single-function boundaries.

\paragraph{LLM-based Methods.}
Recent work has explored large language models for type recovery.
ReSym~\cite{xie2024resym} fine-tunes StarCoder to recover variable names and types and structure field information from decompiled functions, combining its predictions through Prolog-based reasoning.
TypeForge~\cite{wang2025typeforge} synthesizes candidate composite type declarations using Type Flow Graphs and employs LLMs to select among them by comparing the readability of decompiled-code variants.
Both pipelines therefore depend on decompiler output to provide the code context for their LLM-based prediction or selection stages.

\paragraph{Positioning of Our Work.}
Our work combines CodeArt's pre-trained assembly representations with inter-procedural analysis motivated by ByteTR's empirical findings. We encode caller--callee dependencies through CodeArt's existing structured attention mechanism and evaluate their contribution to binary type recovery.

\section{Discussion and Limitations}
\label{sec:discussion}

\paragraph{Integration with Decompilers.}
\toolname can be integrated into existing reverse engineering workflows as a post-processing step.
Given a stripped binary, analysts can use IDA Pro or Ghidra for initial disassembly and then apply \toolname to produce refined type annotations.
The recovered types can be imported back into the decompiler to improve the readability of decompiled code, providing better variable names and structure definitions.

\paragraph{Indirect Calls and Virtual Dispatch.}
Our current approach handles only direct function calls, where the callee's address is statically known.
Indirect calls through function pointers (\eg \texttt{call [rax]}) and virtual method dispatch (\eg C++ vtable calls) present a challenge because the call target cannot be determined statically.
Handling these cases would require integration with call graph recovery techniques or dynamic analysis to resolve indirect call targets.

\paragraph{Scalability.}
The concatenation of caller and callee functions into a single input sequence is bounded by the Transformer's maximum sequence length (512 tokens in our implementation).
For long functions, this may require truncation of the callee.
Future work could explore sparse attention mechanisms or hierarchical encoding to accommodate larger function pairs without truncation.

\paragraph{Inter-Procedural Depth.}
We currently limit inter-procedural analysis to direct caller--callee pairs, spanning one call edge (depth 1 in our terminology). ByteTR~\cite{li2026bytetr} counts function levels instead: its depth 1 is intra-procedural, and its depth 2 covers a direct caller--callee pair. ByteTR reports an 8.46\% precision improvement from one to two levels, followed by an additional 2.08\% improvement from two to three levels. This smaller incremental gain motivates considering the accuracy--cost trade-off when constructing and processing longer caller--callee chains.
Extending to greater depths while maintaining efficiency is a direction for future work.

\section{Conclusion}
\label{sec:conclusion}

We have presented \toolname, a method that embeds inter-procedural analysis into a pre-trained Transformer model for binary type recovery.
By merging caller--callee CFGs and encoding the resulting cross-function data flow edges as graph-pattern entries in the structured attention mask, our approach enables the Transformer to attend to type constraints that propagate across function boundaries.
On the coreutils benchmark at O1, \toolname improves recall by 2.76 percentage points, from 89.00\% for the intra-procedural CodeArt baseline to 91.76\%, corresponding to a 25.1\% reduction in recall error rate ($1-\text{recall}$).

Our results demonstrate that the gap between intra-procedural and inter-procedural type recovery is significant and addressable within the Transformer framework.
Future work could investigate whether the same approach---merging caller--callee CFGs, extracting cross-function data flow edges via reaching definition analysis, and encoding them as graph-pattern attention mask entries---benefits other Transformer-based binary analysis tasks.
For example, \emph{function signature recovery} could use caller argument preparation to help infer parameter counts and types from calling conventions.
For \emph{variable name prediction}, callee context might help disambiguate generic register names by exposing how values are used across function boundaries.
For \emph{vulnerability detection}, cross-function dependencies could provide useful context for identifying use-after-free or buffer overflow patterns involving allocation and use in different functions.
These applications remain possibilities for future evaluation.

\bibliography{references}

\end{document}